\documentclass{article}

\usepackage{arxiv}

\usepackage[utf8x]{inputenc} 
\usepackage[T1]{fontenc}    
\usepackage{hyperref}       
\usepackage{url}            
\usepackage{amsfonts}       
\usepackage{nicefrac}       
\usepackage{microtype}      
\usepackage{graphicx}
\usepackage{amsmath}
\usepackage{amssymb}

\title{On the Stochastic Processes on $7$-Dimensional Spheres}

\author{Nurfa Risha\\
	Departemen Fisika, Fakultas Matematika dan Ilmu Pengetahuan Alam\\
	Universitas Gadjah Mada\\
	Sekip Utara Bulaksumur, Yogyakarta, 55281, Indonesia\\
	\\
	Program Studi Pendidikan Fisika, Fakultas Matematika dan Ilmu Pengetahuan Alam\\
	Universitas Pendidikan Ganesha\\Jalan Udayana No. 10-11, Singaraja 81116, Indonesia.\\
	\texttt{nurfa.risha@undiksha.ac.id}\\
	\And
	Adhitya Ronnie Effendie\\
	Departemen Matematika, Fakultas Matematika dan Ilmu Pengetahuan Alam\\
	Universitas Gadjah Mada\\
	Sekip Utara Bulaksumur, Yogyakarta, 55281, Indonesia\\
	\texttt{adhityaronnie@ugm.ac.id}\\
	\AND
	Muhammad Farchani Rosyid\\
	Departemen Fisika, Fakultas Matematika dan Ilmu Pengetahuan Alam\\
	Universitas Gadjah Mada\\
	Sekip Utara Bulaksumur, Yogyakarta, 55281, Indonesia\\
	\texttt{farchani@ugm.ac.id}\\
}

\begin{document}
	\maketitle
	
	\begin{abstract}
	    We studied isometric stochastic flows of a Stratonovich stochastic differential equation on spheres, i.e. on the standard sphere and Gromoll-Meyer exotic sphere. The standard sphere $S^7_s$ can be constructed as the quotient manifold $\mathrm{Sp}(2, \mathbb{H})/S^3$ with the so-called ${\bullet}$-action of $S^3$, whereas the Gromoll-Meyer exotic sphere $\Sigma^7_{GM}$ as the quotient manifold $\mathrm{Sp}(2, \mathbb{H})/S^3$ with respect to the so-called ${\star}$-action of $S^3$. The Stratonovich stochastic differential equation which describes a continuous-time stochastic process on the standard sphere, is constructed and studied. The corresponding continuous-time stochastic process and its properties on the Gromoll-Meyer exotic sphere can be obtained by constructing a homeomorphism $h: S^7_s\rightarrow \Sigma^7_{GM}$. The corresponding Fokker-Planck equation and entropy rate in the Stratonovich approach is also investigated.
	\end{abstract}
	
	\section{Introduction}
	The microscopic description of the dynamics of a diffusion process on a connected compact differentiable manifold $\mathcal{M}$, for instance, is typically represented by a so-called Stratonovich formulation of stochastic differential equations. Likewise, the dynamic descriptions of the other stochastic processes on differentiable manifolds are also represented in the formulation. The advantage of this formulation is that It\^{o}'s formula appears in the same form as the fundamental theorem of calculus; therefore, stochastic calculus in this formulation takes a more familiar form. The other advantage of Stratonovich formulation, which is more important and relevant to our investigation is that stochastic differential equations on manifolds in this formulation transform consistently under diffeomorphisms between manifolds.

	In general, the laws of physics must be independent of the choice of a coordinate system. This statement means, for instance, that there must be a family of coordinate systems that are compatible with describing space-time. A set of compatible coordinate systems in space-time forms a differential structure. Space-time is an example of a topological space equipped with a differential structure. In a topological space, we may find more than one differential structures. Therefore, from a single topological space, we can construct more than one space-time with different differential structures. If the yielded space-times are not distinguishable in the sense that they are not diffeomorphic, they may lead to inequivalent formulations of the law of physics. The totality of inequivalent differential structures on a topological space is called the exotica on the topological space.

	In mathematics, spheres are topological spaces which are interesting to investigate, and in many branches of physics, they serve for instance as models for configuration spaces of some mechanical systems. In physics, for example, the standard $7$-dimensional sphere $S^7_s$ is particularly interesting in related to supersymmetry breaking and to the work of Witten in which he used it to cancel the global gravitational anomalies in $1985$. Two differential structures on $7$-dimensional sphere $S^7$ are said to be equivalent if there is a diffeomorphism pulling back the differentiable maximal atlas from the second to the first. The connected compact topological space $S^7$ has more than one distinct differential structure that is not equivalent each other in this sense or more precisely there are connected compact $7$-dimensional differentiable manifolds which are homeomorphic but not diffeomorphic to the standard seven-sphere $S^7_s$. The differential structure on the standard sphere $S^7_s$ is called a standard differential structure, while the differential structures that are not equivalent to the standard one are called exotic differential structures. The topological space $S^7$ equipped with an exotic differential structure or a seven-dimensional differentiable manifold which is homeomorphic but not diffeomorphic to standard seven-dimensional sphere $S^7_s$ is called exotic $7$-sphere.

	Recently, Gromoll and Meyer have constructed a free action of $S^3$ on $\mathrm{Sp}(2, \mathbb{H})$ which preserves the bi-invariant metric. The quotient is an exotic $7$-sphere. The first time where biquotients were considered in geometry, was in \cite{Gromoll}, where it was shown that an exotic $7$-sphere admits non-negative curvature. The Gromoll-Meyer $7$-sphere is the only exotic sphere which can be written as a biquotient. The concept of a biqoutient $\mathrm{Sp}(2, \mathbb{H})//\mathrm{Sp}(1, \mathbb{H})$, is an exotic $7$-sphere, which produced the first example of an exotic sphere with non-negative curvature. The non-negatively curved manifold $\mathrm{Sp}(2, \mathbb{H})//\mathrm{Sp}(1, \mathbb{H})$ is homeomorphic, but not diffeomorphic, to $S^7_s$. The Gromoll-Meyer sphere $\mathrm{Sp}(2, \mathbb{H})//\mathrm{Sp}(1, \mathbb{H})$ is obtained by letting $\mathrm{Sp}(1, \mathbb{H})$ act via diag-$(q, q)$ on the left, and diag-$(q, 1)$ on the right.

	The properties of the Gromoll-Meyer exotic sphere $\Sigma^7_{GM}$ related to the isometric stochastic flow are to be studied in this article. The properties of an isometric stochastic flow of the Stratonovich stochastic differential equation on the $7$-dimensional spheres will be examined. The seven spheres which will be considered here are the standard sphere $S^7_s$ and the Gromoll-Meyer exotic sphere $\Sigma^7_{GM}$. In this study, $\Sigma^7_{GM}$ is understood to be the $7$-dimensional sphere equipped with a Gromoll-Meyer (exotic) differential structure. The standard sphere $S^7_s$ can be regarded as a quotient manifold $\mathrm{Sp}(2, \mathbb{H})/S^3$ through ${\bullet}$-action by $S^3$ and $\Sigma^7_{GM}$ regarded as quotient manifold $\mathrm{Sp}(2, \mathbb{H})/S^3$ but with respect to the ${\star}$-action of $S^3$. A Stratonovich stochastic differential equation which describes the continuous stochastic process (i.e., the stochastic process which is parameterized with continuous-time) on the standard sphere, is constructed and examined. Regarding the continuous stochastic process formulated in $\Sigma^7_{GM}$, which is obtained by constructing a homeomorphism $h: S^7_s\rightarrow \Sigma^7_{GM}$. The Fokker-Planck equation and entropy rate in the Stratonovich approach are also studied in both spheres.
	
	\section{Preliminaries}
	\subsection{Continuous stochastic process}
	Let $(\Omega, \mathcal{F}, \mathbb{P})$ be a probability space, $(E, \mathcal{E})$ a measurable space, and $T$ a set. A mapping $Z:T\times (\Omega, \mathcal{F})\rightarrow (E, \mathcal{E})$ is called a stochastic process with phase space $(E, \mathcal{E})$ if the measurable maps $ (Z_t)_{t\in T}:(\Omega, \mathcal{F})\rightarrow (E, \mathcal{E})$, where $ (Z_t)_{t\in T}(\omega)=z_t$ for every $\omega \in \Omega$, is a random field taking values in a measurable space $(E, \mathcal{E})$. The sample paths of $Z$ are the mappings $Z(\omega):T\rightarrow (E, \mathcal{E}); t\mapsto (Z_t)_{t\in T}(\omega)$ obtained by fixing $\omega\in \Omega$. The sample paths of $Z$ thus form a family of mappings from $T$ into $(E, \mathcal{E})$ indexed by $\omega\in \Omega$. A stochastic process $Z=\{(Z_t)_{t\in T}(\omega)\}$ is said to be continuous, right-continuous, or left-continuous if for $\mathbb{P}$-almost all $\omega$, the sample path of $\omega$ is continuous, right-continuous, or left-continuous, respectively. A continuous-time stochastic process $Z=\{(Z_t)_{t\in T}(\omega)\} $ is said to have continuous sample paths, if $t\mapsto (Z_t)_{t\in T}(\omega)$ is continuous for all $\omega\in\Omega$ \cite{Emery}.

	The Brownian motion $B=\{(B_t)_{t\in T}(\omega)\}$ with $T=\{0\leq t <\infty\}$ is a continuous martingale, and a continuous local martingale can be (random) time changed to be a Brownian motion. Therefore, the study of Brownian motion has much to say about the sample path properties of much more general classes of processes, including continuous martingales and diffusion process. 
	\subsection{Stochastic differential equation}
	In most applications, the phase space of a continuous-time stochastic process can be regarded as $(\mathbb{R}^{d}, \mathcal{B})$, i.e., the $d$-dimensional Euclidean space equipped with the $\sigma$-algebra of Borel sets concerning the natural topology in $\mathbb{R}^{d}$. However, many applications are demanding another kind of measurable spaces $(E, \mathcal{E})$ to play the role of phase space. In this study, we consider manifolds which are homeomorphic to $S^7_s$ as the phase space of a continuous-time stochastic process.

	The significant advantage of the Stratonovich stochastic integral is that it obeys the usual transformation rules of calculus. Stratonovich motivation was to achieve a rigorous treatment of the stochastic differential equation, which governs the diffusion processes of A. N. Kolmogorov. For this reason, Stratonovich formula is often used to formulate stochastic differential equations on manifolds such as a circle or sphere. Stochastic processes defined by Stratonovich integrals over a varying time interval does not, however, satisfy the powerful martingale properties of their It\^{o} integral counterparts as we have already mentioned.

	Let $\mathcal{M}$ be an $(d-1)$-dimensional differentiable manifold embedded in $\mathbb{R}^{d}$. The usual continuous-time stochastic process on the manifold $\mathcal{M}$ is described by the following Stratonovich stochastic differential equation of the following form \cite{Kunita}
	\begin{equation}
	dz_t=U_0(z_t) dt+ \sum _{i=1}^{d-1} U_i(z_t)\circ dB_t^i(\omega),
	\label{sde01}
	\end{equation}
	where $\{B_t^1(\omega), B_t^2(\omega), \cdots, B_t^{(d-1)} (\omega)\}$ are standard $(d-1)$-dimensional Brownian motions constructed on a filtered probability space $(\Omega, \mathcal{F}, (\mathcal{F}_t)_{t\geq 0}, \mathbb{P})$, $\circ dB^i_t(\omega)$ indicates that the integral involved being Stratonovich, and $\{U_0, U_1, \cdots,$ $ U_{(d-1)}\}$ is a $C^\infty$-vector field on the manifold $\mathcal{M}$. Equation \eqref{sde01} has a solution flow $g_{s, t}:\Omega\times \mathcal{M}\rightarrow\mathcal{M}$, consisting of random smooth diffeomorphisms of $\mathcal{M}$, continuous in $t$.

	The last term of equation \eqref{sde01} can be regarded as a noise or a random disturbance adjoined to the ordinary differential equation $dz_t=U_0(z_t) dt$. Here the continuous sample paths $t\mapsto B^i_t(\omega)$ are not functions of bounded variations with respect to $t$ a.s, so that $dB^i_t(\omega)$ cannot be defined as the Stieltjes integral. Nevertheless the last integrals are well defined for almost all samples if $z_t$ is adapted, i.e., for any $t$, $z_t$ is independent of the future Brownian motion $B_s(\omega)-B_t(\omega)$, $s\geq t$. The Stratonovich stochastic differential equation \eqref{sde01} can also be written more concisely as $dz_t=\sum_{i=0}^{(d-1)} U_i(z_t)\circ dB^i_t(\omega)$, where $B_t^0(\omega)=t$.

	Let $z^{(n)}=(z_1, z_2, \cdots, z_n)$ be $n$-point in $\mathcal{M}$ or an element of $\mathcal{M}$. Set $g_{s, t}(\omega, z^{(n)})=(g_{s, t}(\omega, z_1)$, $g_{s, t}(\omega, z_2), \cdots, g_{s, t}(\omega, z_n), t\geq s)$. Then for each fixed $s\geq 0$ and $z^{(n)}$, $g_{s, t}(\omega, z^{(n)}), t\in [s, T]$ is a continuous-time stochastic process with values in $\mathcal{M}$ starting at $z^{(n)}$ at time $s$. It is called an $n$-point motion of the flow $g_{s, t}(\omega)$. Stochastic flows on manifolds are defined similarly to those on Euclidean space.

	Let $\mathcal{M}$ be a smooth $(d-1)$-dimensional manifold and $G$ be a Lie group of diffeomorphisms $\mathcal{M}\rightarrow\mathcal{M}$. Suppose $\phi_t$ is a diffusion on $G$ with $\phi_0=I$, $I$ the identity of $G$. $\phi_t$ will be called a $G$-valued stochastic flow on $\mathcal{M}$. For any $x\in\mathcal{M}$, $\phi_t(x)$ is a stochastic process on $\mathcal{M}$ with $\phi_0(x)=x$ and will be called the one-point motion of $\phi_t$ starting from $x$. There has been an extensive effort over the past few years to study the stochastic flow of smooth random fields, $g$, from a general Lie group to Riemannian manifold $\mathcal{M}$ \cite{Liao}. 
	
	\section{Isometric stochastic flows and Gromoll-Meyer exotic sphere}
	\subsection{Isometric stochastic flows}
    If $\mathcal{M}$ is endowed with a Riemannian metric $\textbf{g}$, an action $l:G\times \mathcal{M}\rightarrow\mathcal{M}$ of a group $G$ on $(\mathcal{M}, \textbf{g})$ is said to be isometric (or by isometries) if $\l^g: \mathcal{M}\rightarrow\mathcal{M}$ is an isometry of $(\mathcal{M}, \textbf{g})$ for all $g\in G$. In this case, the metric $\textbf{g}$ is said to be $G$-invariant, and $\l^G$ can be identified with a subgroup of $Iso(\mathcal{M}, \textbf{g})$. A $G$-valued stochastic flow will be called an isometric stochastic flow, if and only if $\mathcal{M}$ is a Riemannian manifold and $G$ is a Lie group of isometries. In order for the stochastic flow generated by equation \eqref{sde01} to be isometric (the equation \eqref{sde01} is contained in the isometry group $G$ of $\mathcal{M}$) and associated to Brownian motion on $\mathcal{M}$, the vector fields $U_i$ in \eqref{sde01} must be Killing vector fields and the sum of $U_i U_i+U_0$ is the Laplace-Beltrami operator on $\mathcal{M}$. Hence, if there are Killing vector fields $U_i, i=0, 1, \cdots, (d-1)$, on $\mathcal{M}$ such that they form the Laplace-Beltrami operator, then there is an isometric stochastic flow whose one point motion is a Brownian motion on $\mathcal{M}$ \cite{Liao}. Therefore, a mapping $g_{s, t}(\omega, \cdot): x\in \mathcal{M}\mapsto g_{s, t}(\omega, x)=x_t\in \mathcal{M}$ is an isometric stochastic flows whose one point motion is a Brownian motion on $\mathcal{M}$, i.e., a diffusion process generated by equation \eqref{sde01} \cite{Kunita}. The map $g_{s, t}(\omega)$ will be referred to as the isometric stochastic flows associated to \eqref{sde01}. Furthermore, the isometric stochastic flows act naturally on Killing vector fields.

We consider the standard $7$-sphere, isometrically embedded as the unit sphere in a $8$-dimensional Euclidean vector space $(V,\textbf{g})$: $S^7_s=\{x\in V:\textbf{g}(x,x) =1\}\subset V$. Let $\mathrm{SO}(8, \mathbb{R})$ be a special orthogonal group acting on $S^7_s\subset \mathbb{R}^8$. The isometric action of $SO(8,\mathbb{R})$ on $\mathbb{R}^8$ is transitive on the unit sphere. The solution of a Stratonovich stochastic differential equation on $SO(8,\mathbb{R})$ is an $SO(8,\mathbb{R})$-valued isometric stochastic flow $g_{s, t}(\omega)$ on $S^7_s$. In fact, $U_0=0$ and $U_1, U_2, \cdots, U_{7}$ is vector fields on $S^7_s$ of unit speed rotations determined by $2$-dimensional coordinate planes in $\mathbb{R}^8$. The vector fields on $S^7_s$ are the Killing vector field that forms Laplace-Beltrami operator of $S^7_s$. Therefore, there is an isometric stochastic flow $g_{s, t}(\omega)$ on $S^7_s$ whose one point motion is a Brownian motion. Hence, the motion of $S^7_s$ can be described by an isometric stochastic flow $g_{s, t}(\omega)$ on $S^7_s$. The motion of a fixed point on $S^7_s$ is a Brownian motion $x_t$ on $S^7_s$, will be called the one point motion $x_t$ of $g_{s, t}(\omega)$.

The set of maps $g_{s,t}(\omega)$ is called an \textit{isometric stochastic flow of homeomorphisms} if there exists a null set $N$ of $\Omega$ so that for any $\omega\in N^c$, the set of continuous maps $\{g_{s,t}(\omega):s,t\in [0, T]\}$ defines an isometric stochastic flow of homeomorphisms, i.e., it satisfies the following properties \cite{Kunita}
\begin{enumerate}
	\item [(1)]$g_{s,u}(\omega)=g_{t,u}(\omega)\circ g_{s,t}(\omega)$ (cocycle property) holds for all $s, t, u$ whenever $0\leq s\leq t\leq u$ for all $\omega\in\Omega$, with no exceptions, where $\circ$ denotes the composition of maps.
	
	If we use the notation $g_{s, t}(\omega, z)$ instead of $g_{s, t}(\omega)(z)$, for $z\in S^7_s$, then (1) becomes:
	\begin{equation}
	g_{t, u}(g_{s,t}(\omega, z), \omega)=g_{s, u}(\omega, z), z\in S^7_s, \omega\in \Omega.
	\label{flow03}
	\end{equation}
	Loosely speaking, each random field $g_{s, t}$ is a random transformation of $S^7_s$, and \eqref{flow03} is a kind of semigroup property, often called the flow property. A Gaussian random field is a random field where any finite collection of elements $g_{s, t}(z_1),$ $\cdots,$ $g_{s, t}(z_8)$ has a multidimensional Gaussian distribution.
	\item [(2)] $g_{s,s}(\omega)$ $=$ identity map for all $s\in [0, T]$ and all $\omega\in\Omega$.
	\item [(3)] the map $g_{s,t}(\omega, \cdot): S^7_s\rightarrow S^7_s$ is an onto homeomorphism for all $s,t\in [0, T]$.

	This $g_{s, t}$ is called an isometric stochastic flow of homeomorphisms generated by the Killing vector field $U_i, i=0, 1, \cdots, (d-1)$.
\end{enumerate}
Further, if $g_{s,t}(\omega)$ satisfies 
\begin{enumerate}
	\item[(4)] $g_{s,t}(\omega, z)$ is $k$-times differentiable with respect to $z$ for all $s,t$ and the derivatives are continuous in $(s, t, z)$,
\end{enumerate}
Then it is called an \textit{isometric stochastic flow of $C^k$ diffeomorphisms}.

Let $g_{s,t}(\omega)^{-1}$ be the inverse map of $g_{s,t}(\omega)$. Then the conditions (1) and (2) imply $g_{t, s}(\omega)=g_{s,t}(\omega)^{-1}$, $s\leq t$. This fact and the condition (3) show that $g_{s,t}(\omega)^{-1}(z)$ is also continuous in $(s, t, z)$. The condition (4) implies that $g_{s,t}(\omega)^{-1}(z)$ is $k$-times continuously differentiable with respect to $z$. Hence $g_{s,t}(\omega, \cdot): S^7_s \rightarrow S^7_s$ is actually a $C^k$-diffeomorphism for all $s,t$ if (4) is satisfied. We can regard $g_{s,t}(\omega)^{-1}(z)$ as a random field with parameter $(s, t, z)$, and denotes it as $g_{s, t}^{-1}(\omega, z)$. Thus $g_{s, t}^{-1}(\omega, z)=g_{t, s}(\omega, z)$ holds for all $s, t, z$ a.s. An isometric stochastic flow of homeomorphisms can be considered as a continuous random field with values in $\mathrm{SO}(8, \mathbb{R})$ satisfying the flow properties $(1)$ and $(2)$. We will call it a \textit{continuous isometric stochastic flow with values in $\mathrm{SO}(8, \mathbb{R})$} \cite{Kunita}.

Let $g_t (\omega), t\geq 0$ be a diffusion process with values in $\mathrm{SO}(8, \mathbb{R})$ such that $g_0(\omega)=e_{\mathrm{SO}(8, \mathbb{R})}$ a.s. Set $g_{s, t}(\omega)=g_t(\omega)\circ g_s(\omega)^{-1}$ where $g_s(\omega)^{-1}$ is the inverse of $g_s(\omega)$. Then it is an isometric stochastic flow with values in $\mathrm{SO}(8, \mathbb{R})$. Conversely if $g_{s, t}(\omega)$ is an isometric stochastic flow with values in $\mathrm{SO}(8, \mathbb{R})$, then $g_t(\omega)\equiv g_{0, t}(\omega)$ is a diffusion process with values in $\mathrm{SO}(8, \mathbb{R})$ such that $g_0(\omega)=e_{\mathrm{SO}(8, \mathbb{R})}$ a.s. It satisfies $g_{s, t}(\omega)=g_t(\omega)\circ g_s(\omega)^{-1}$ by the flow property $(1)$. Therefore a continuous isometric stochastic flow with values in $\mathrm{SO}(8, \mathbb{R})$ is equivalent to a diffusion process $g_t(\omega)$ with values in $\mathrm{SO}(8, \mathbb{R})$ such that $g_0(\omega)=e_{\mathrm{SO}(8, \mathbb{R})}$ a.s.

However, $S^7_s$ is a connected compact Riemannian manifold and $\mathrm{SO}(8, \mathbb{R})$ is set of all the Lie transformation group of isometries on $S^7_s$. A diffusion process on $S^7_s$ can due to an isometric stochastic flow $g_{s, t}(\omega)$ of global diffeomorphisms on a Lie transformation group $\mathrm{SO}(8, \mathbb{R})$ acting naturally on $S^7_s$. Such continuous-time stochastic process on $S^7_s$ is the form one point motions $z_t$ on $S^7_s$ \cite{Liao}. The isometric stochastic flows on $S^7_s$ associated to such one point motion is called isometric stochastic flows on $S^7_s$ induced by left $\mathrm{SO}(8, \mathbb{R})$-invariant diffusion process $g_{s, t}(\omega)$.

\subsection{Gromoll-Meyer exotic sphere}
John Milnor found the first examples of exotic $7$-sphere in $1956$ among the principal $S^3$-bundles over $S^4$. Milnor's construction makes use of a generalization of the Hopf fibrations. Gromoll and Meyer \cite{Gromoll} constructed an exotic sphere $\Sigma^7_{GM}$, as a bi-quotient of the compact Lie transformation group $\mathrm{Sp}(2, \mathbb{H})$ of quaternionic unitary matrices by an explicit $S^3$ action. The bi-invariant metric of $\mathrm{Sp}(2, \mathbb{H})$ is subduced onto $\Sigma^7_{GM}$ to provide $\Sigma^7_{GM}$ automatically with a Riemannian metric of non-negative sectional curvature ($K\geq 0$). Every bi-quotient of the compact Lie transformation group of quaternionic unitary matrices has a Riemannian metric of non-negative sectional curvature and thereby $\Sigma^7_{GM}$ is the first exotic sphere with non-negative sectional curvature.

In 2002 Totaro \cite{Totaro} and independently Kapovitch and Ziller \cite{Kapovitch} showed that $\Sigma^7_{GM}$ is the only exotic sphere that can be expressed as a bi-quotient of a compact Lie transformation group. Note that $\Sigma^7_{GM}$ can be regarded as the basic example of a bi-quotient in Riemannian geometry and, simultaneously as the core example of an exotic sphere \cite{Totaro}. In fact, $\Sigma^7_{GM}$ is diffeomorphic to the Milnor exotic sphere $\Sigma^7_{2, -1}$.

There are two free isometric actions of $S^3$ on $\mathrm{Sp}(2, \mathbb{H})$ (which are isometric for many metrics and connections) that plays a central role in the rest of the paper. Both of these actions foliate $\mathrm{Sp}(2, \mathbb{H})$ by $S^3$-orbits in two different ways. The first action is the standard action given as follows : if $q\in S^3$ and $Q\in \mathrm{Sp}(2, \mathbb{H})$ \cite{Gromoll},
\begin{eqnarray}
S^3\times \mathrm{Sp}(2, \mathbb{H})&\rightarrow& \mathrm{Sp}(2, \mathbb{H}):\nonumber\\
(q, Q)\mapsto q\bullet Q&=&Q\begin{pmatrix}
1&0\\
0&\bar{q}
\end{pmatrix}=\begin{pmatrix}
a&b\\
c&d
\end{pmatrix}\begin{pmatrix}
1&0\\
0&\bar{q}
\end{pmatrix}=\begin{pmatrix}
a&b\bar{q}\\
c&d\bar{q}
\end{pmatrix},
\label{dott04}
\end{eqnarray}
where 
\begin{equation}
\mathrm{Sp}(2, \mathbb{H})=\bigg\{Q=\begin{pmatrix}
a&b\\
c&d
\end{pmatrix}\in S^7_s\times S^7_s\bigg|\bar{b}a+\bar{d}c=0\bigg\}.
\label{Sp(2, H)05}
\end{equation}
The action therefore leads to a principal fibration $S^3\rightarrow \mathrm{Sp}(2, \mathbb{H})\rightarrow S^7_s$ called $\bullet$-fibration. The orbit space of the standard $\bullet$-action can be naturally identified with $S^7_s\subset \mathbb{H}^2$ by restricting a matrix in $\mathrm{Sp}(2, \mathbb{H})$ to its first column \cite{Speranca}. Let $\pi_{\bullet }$ be the projection map of the above principal bundle with the base space being the $S^7_s$. The projection is therefore given by 
\begin{eqnarray}
\pi_{\bullet }:\mathrm{Sp}(2, \mathbb{H})&\rightarrow& S^7_s\nonumber\\
Q=\begin{pmatrix}
a&b\\
c&d
\end{pmatrix}&\mapsto& (a, c)\in S^7_s. 
\label{map06}
\end{eqnarray}
The second one is the Gromoll-Meyer exotic action given by \cite{Gromoll}
\begin{eqnarray}
S^3\times \mathrm{Sp}(2, \mathbb{H})&\rightarrow& \mathrm{Sp}(2, \mathbb{H}):\nonumber\\
(q, Q)\mapsto q\star Q&=&qQ\begin{pmatrix}
\bar{q}&0\\
0&1
\end{pmatrix}\nonumber\\
&=&\begin{pmatrix}
q&0\\
0&q
\end{pmatrix}\begin{pmatrix}
a&b\\
c&d
\end{pmatrix}\begin{pmatrix}
\bar{q}&0\\
0&1
\end{pmatrix}=\begin{pmatrix}
qa\bar{q}&qb\\
qc\bar{q}&qd
\end{pmatrix}.
\label{starr07}
\end{eqnarray}
The second action leads to a principal fibration $S^3\rightarrow \mathrm{Sp}(2, \mathbb{H})\rightarrow \Sigma^7_{GM}$ called $\star$-fibration. The base manifold of this fibration is the well-known $\Sigma^7_{GM}$.

	\section{Main Results and Discussion}
	\subsection{Stochastic processes on $S^7_s$}
    As we have mentioned before that the class 
\begin{equation}
\Bigg[\begin{pmatrix}
a&b\\
c&d
\end{pmatrix}\Bigg]^\bullet \in S^7_s,
\label{S^708}
\end{equation}
can be identified with $(b,d)\in \mathbb{H}\times \mathbb{H}$ satisfying $b\bar{b}+d\bar{d}=1$,
where $b=b_0+b_1\mathrm{i}+b_2 \mathrm{j}+b_3 \mathrm{k}$ and $d=d_0+d_1\mathrm{i}+d_2\mathrm{j}+d_3 \mathrm{k}$. Now consider the unit sphere $S^7_s$ as the submanifold of $\mathbb{R}^8$, i.e., a $S^{7}$ is given as a set of points in $\mathbb{R}^8$ whose coordinates $(z^1,z^2,\cdots, z^8)$ satisfy $\sum_{i=1}^{8} (z^i)^2=1$, namely
\begin{equation}
S^7_s=\bigg\{z=(z^1, z^2, \cdots, z^8)\in \mathbb{R}^8\bigg|\Arrowvert z\Arrowvert^2_E=1\bigg\}.
\label{S^709}
\end{equation}
Let $(b, d)$ be any point on $S^7_s$. The point $(b,d)$ can be identified for instance with a point $z=(z^1,z^2\cdots, z^8)\in \mathbb{R}^8$, as follows
\begin{eqnarray}
z^1&=&b_0 \qquad z^2=d_0\nonumber\\
z^3&=&b_1\qquad z^4=d_1\nonumber\\
z^5&=&b_2\qquad z^6=d_2\nonumber\\
z^7&=&b_3\qquad z^8=d_3.
\label{identification10}
\end{eqnarray}
Since $b\bar{b}+d\bar{d}=1$,
\begin{eqnarray}
&&(z^1+z^3\mathrm{i}+z^5\mathrm{j}+z^7\mathrm{k})(z^1-z^3\mathrm{i}-z^5\mathrm{j}-z^7\mathrm{k}) +(z^2+z^4\mathrm{i}+z^6\mathrm{j}+z^8\mathrm{k})(z^2-z^4\mathrm{i}-z^6\mathrm{j}-z^8\mathrm{k})\nonumber\\
&&=(z^1)^2+(z^2)^2+(z^3)^2+(z^4)^2+(z^5)^2+(z^6)^2+(z^7)^2+(z^8)^2\nonumber\\
&&=1.
\label{identification11}
\end{eqnarray}
Therefore, the point $z=(z^1, z^2, z^3, \cdots, z^8)\in \mathbb{R}^8$ which is identified with $(b,d)$ is in $S^7_s$ $\subset$ $\mathbb{R}^8$.

\subsubsection{\textbf{Orthonormal vector field on $S^7_s$}}
An interesting fact of $S^7_s$ is that we can find a globally defined frame on it, i.e. globally nonvanishing vector fields whose values at every point form a basis for the tangent space. From short observation, it is clear that the tangent space at a point $z\in S^7_s$ is the $7$-dimensional subspace of $\mathbb{R}^{8}$ consisting of all vectors which are perpendicular to $z$. In particular every vector field on $S^7_s$ can be represented by a continuous function $v:S^7_s\rightarrow\mathbb{R}^8$ such that $v(z)\hspace{.1cm}\bot\hspace{.1cm} z$ (or $\langle z, v(z)\rangle=0$) for all $z\in S^7_s$. In this way, the linear structure of vector fields become even more apparent. The idea is to assign to each point $z\in S^7_s$ a vector $v(z)$ tangent to $z$ in a smooth way. However, by assumption $v(z)$ is nonvanishing, so we can normalize such that $\| v(z)\|$. Therefore, there exist seven nonvanishing linearly independent vector fields on $S^7_s$.

For every point $z=(z^1, z^2, \cdots, z^8)\in S^7_s$, the following vector fields on $S^7_s$ are nonvanishing and linearly independent \cite{Furutami}:
\begin{eqnarray}
U_1&=&-z^2\partial_{1}+z^1\partial_{2}-z^4\partial_{3}+z^3\partial_{4}-z^6\partial_{5}+z^5\partial_{6}-z^8\partial_{7}+z^7\partial_{8}\nonumber\\
&=&\bigg(z^1\frac{\partial}{\partial z^2}-z^2\frac{\partial}{\partial z^1}\bigg)+\bigg(z^3\frac{\partial}{\partial z^4}-z^4\frac{\partial}{\partial z^3}\bigg)+\bigg(z^5\frac{\partial}{\partial z^6}-z^6\frac{\partial}{\partial z^5}\bigg)+\bigg(z^7\frac{\partial}{\partial z^8}-z^8\frac{\partial}{\partial z^7}\bigg)\nonumber \\
&=&U_{12}+U_{34}+U_{56}+U_{78}\nonumber
\end{eqnarray}
\begin{eqnarray}
U_2&=&-z^3\partial_{1}+z^4\partial_{2}+z^1\partial_{3}-z^2\partial_{4}+z^7\partial_{5}-z^8\partial_{6}-z^5\partial_{7}+z^6\partial_{8}\nonumber\\
&=&\bigg(z^1\frac{\partial}{\partial z^3}-z^3\frac{\partial}{\partial z^1}\bigg)-\bigg(z^2\frac{\partial}{\partial z^4}-z^4\frac{\partial}{\partial z^2}\bigg)-\bigg(z^5\frac{\partial}{\partial z^7}-z^7\frac{\partial}{\partial z^5}\bigg)+\bigg(z^6\frac{\partial}{\partial z^8}-z^8\frac{\partial}{\partial z^6}\bigg)\nonumber \\
&=&U_{13}-U_{24}-U_{57}+U_{68}\nonumber
\end{eqnarray}
\begin{eqnarray}
U_3&=&-z^4\partial_{1}-z^3\partial_{2}+z^2\partial_{3}+z^1\partial_{4}-z^8\partial_{5}-z^7\partial_{6}+z^6\partial_{7}+z^5\partial_{8}\nonumber\\
&=&\bigg(z^1\frac{\partial}{\partial z^4}-z^4\frac{\partial}{\partial z^1}\bigg)+\bigg(z^2\frac{\partial}{\partial z^3}-z^3\frac{\partial}{\partial z^2}\bigg)+\bigg(z^5\frac{\partial}{\partial z^8}-z^8\frac{\partial}{\partial z^5}\bigg)+\bigg(z^6\frac{\partial}{\partial z^7}-z^7\frac{\partial}{\partial z^6}\bigg)\nonumber \\
&=&U_{14}+U_{23}+U_{58}+U_{67}\nonumber
\end{eqnarray}
\begin{eqnarray}
U_4&=&-z^5\partial_{1}+z^6\partial_{2}-z^7\partial_{3}+z^8\partial_{4}+z^1\partial_{5}-z^2\partial_{6}+z^3\partial_{7}-z^4\partial_{8}\nonumber\\
&=&\bigg(z^1\frac{\partial}{\partial z^5}-z^5\frac{\partial}{\partial z^1}\bigg)-\bigg(z^2\frac{\partial}{\partial z^6}-z^6\frac{\partial}{\partial z^2}\bigg)+\bigg(z^3\frac{\partial}{\partial z^7}-z^7\frac{\partial}{\partial z^3}\bigg)-\bigg(z^4\frac{\partial}{\partial z^8}-z^8\frac{\partial}{\partial z^4}\bigg)\nonumber \\
&=&U_{15}-U_{26}+U_{37}-U_{48}\nonumber
\end{eqnarray}
\begin{eqnarray}
U_5&=&-z^6\partial_{1}-z^5\partial_{2}+z^8\partial_{3}+z^7\partial_{4}+z^2\partial_{5}+z^1\partial_{6}-z^4\partial_{7}-z^3\partial_{8}\nonumber\\
&=&\bigg(z^1\frac{\partial}{\partial z^6}-z^6\frac{\partial}{\partial z^1}\bigg)+\bigg(z^2\frac{\partial}{\partial z^5}-z^5\frac{\partial}{\partial z^2}\bigg)-\bigg(z^3\frac{\partial}{\partial z^8}-z^8\frac{\partial}{\partial z^3}\bigg)-\bigg(z^4\frac{\partial}{\partial z^7}-z^7\frac{\partial}{\partial z^4}\bigg)\nonumber \\
&=&U_{16}+U_{25}-U_{38}-U_{47}\nonumber
\end{eqnarray}
\begin{eqnarray}
U_6&=&-z^7\partial_{1}+z^8\partial_{2}+z^5\partial_{3}-z^6\partial_{4}-z^3\partial_{5}+z^4\partial_{6}+z^1\partial_{7}-z^2\partial_{8}\nonumber\\
&=&\bigg(z^1\frac{\partial}{\partial z^7}-z^7\frac{\partial}{\partial z^1}\bigg)-\bigg(z^2\frac{\partial}{\partial z^8}-z^8\frac{\partial}{\partial z^2}\bigg)-\bigg(z^3\frac{\partial}{\partial z^5}-z^5\frac{\partial}{\partial z^3}\bigg)+\bigg(z^4\frac{\partial}{\partial z^6}-z^6\frac{\partial}{\partial z^4}\bigg)\nonumber \\
&=&U_{17}-U_{28}-U_{35}+U_{46}\nonumber
\end{eqnarray}
\begin{eqnarray}
U_7&=&-z^8\partial_{1}-z^7\partial_{2}-z^6\partial_{3}-z^5\partial_{4}+z^4\partial_{5}+z^3\partial_{6}+z^2\partial_{7}+z^1\partial_{8}\nonumber\\
&=&\bigg(z^1\frac{\partial}{\partial z^8}-z^8\frac{\partial}{\partial z^1}\bigg)+\bigg(z^2\frac{\partial}{\partial z^7}-z^7\frac{\partial}{\partial z^2}\bigg)+\bigg(z^3\frac{\partial}{\partial z^6}-z^6\frac{\partial}{\partial z^3}\bigg)+\bigg(z^4\frac{\partial}{\partial z^5}-z^5\frac{\partial}{\partial z^4}\bigg)\nonumber\\
&=&U_{18}+U_{27}+U_{36}+U_{45}.
\label{vector fields12}
\end{eqnarray}
It can be shown that the frame $\{U_1, U_2, \cdots, U_7\}$ is left invariant and orthonormal at each point $z\in S^7_s$, with respect to the Euclidean metric in $\mathbb{R}^8$.

\subsubsection{\textbf{Stratonovich stochastic differential equations on $S^7_s$}}

The Stratonovich stochastic differential equation on $S^7_s$ that we will construct in this part is more general than equation \eqref{sde01}, in the sense that we will have as integrator not only a Brownian motion but also an arbitrary continuous semimartingale $W$. Continuous semimartingales are a useful (and widely used) tool for two non-independent reasons: many processes that can be defined explicitly are continuous semimartingales (for instance, diffusion processes); many operations applied to continuous semimartingales yield continuous semimartingales. The space of continuous semimartingales is an appropriate frame for stochastic calculus \cite{Emery}.

A Stratonovich stochastic differential equation on $S^7_s$ can be defined by $(d+1)$-differentiable linearly independent vector fields $\{V_0, V_1, V_2, \cdots, V_d\}$ on $S^7_s$ and a $\mathbb{R}^d$-valued continuous semimartingale $W_t=\{W_t, t\geq 0\}$ and it is viewed as a column of $d$-real valued continuous semimartingales: $W_t=(W_t^1, W_t^2, \cdots, W_t^d)^\dagger$. Each $V_i, i=1, 2, \cdots, d$ can be written as a function $V_i:S^7_s\rightarrow\mathbb{R}^8$ with $V_i(\vec{r})\cdot \vec{r}=0, \forall\; \vec{r}\in\mathbb{R}^8$ with $|\vec{r}|=1$ so that $V=(V_1, V_2, \cdots, V_d)$ is an $(8\times d)$-matrix whose columns are orthogonal and have norm 1. We write the Stratonovich stochastic differential equation generated by $\{V_0, V_1, V_2, \cdots, V_d\}$ and $W_t$ on $S^7_s$ as follows
\begin{equation}
dz_t=V_{\alpha }(z_t)\circ dW_{t}^{\alpha }+V_0(z_t)dt.
\label{sde13}
\end{equation}
Equation \eqref{sde13} can be written explicitly as
\begin{eqnarray}
dz_t^1&=&v_{11}\circ dW_t^1+v_{12}\circ dW_t^2+\cdots+v_{1d}\circ dW_t^d+v_{10}\:dt\nonumber\\
\vdots\;\;& &\hspace{2.0cm}\vdots \hspace{2.0cm}\vdots \nonumber\\
dz_t^8&=&v_{81}\circ dW_t^1+v_{82}\circ dW_t^2+\cdots+v_{8d}\circ dW_t^d+v_{80}\:dt,
\label{sde14}
\end{eqnarray}
where $v_{1\alpha }, v_{2\alpha }, \cdots, v_{8\alpha }$ is the components of $V_{\alpha }$ ($\alpha = 1,2,\cdots, d$), i.e. they satisfy $z^1\cdot v_{1\alpha }(z)+z^2\cdot v_{2\alpha }(z)+\cdots+z^8\cdot v_{8\alpha }(z)=0$ for all $z\in S^7_s$. In fact, stochastic flow generated by equations \eqref{sde13} and \eqref{sde14} is in general not isometric, unless $V_0, V_1, V_2, \cdots,$ and $V_d$ are Killing vector fields.

All the vector fields containing in the frame we have discussed earlier, i.e.,\- $\{U_1, U_2,\- U_3, \cdots, U_7\}$, are Killing vector fields on $S^7_s$, since the vector fields of the form
\begin{equation}
z^i\frac{\partial }{\partial z^j}-z^j\frac{\partial}{\partial z^i}, \quad i\neq j
\label{Killing vector field}
\end{equation}
are Killing and the fact that the Lie derivative of the metric $g$ satisfies $\mathcal{L}_{V+W}\:g=\mathcal{L}_V\:g+\mathcal{L}_W\:g$ for arbitrary vector fields $V$ and $W$. For those vector fields in the frame, we obtain the following systems of Stratonovich stochastic differential equations
\begin{equation}
d\bar{z}_{{\mu }\:t}=U_{\mu }(\bar{z}_t)\circ dW_t,\hspace{1.0cm}(\mu =1,2, \cdots, 7)
\label{sde16}
\end{equation}
or
\begin{eqnarray}
d\bar{z}_{1\:t}=\begin{bmatrix}
-z^2_t\circ dW_t\\
z^1_t \circ dW_t \\ 
-z^4_t\circ dW_t \\
z^3_t\circ dW_t \\ 
-z^6_t\circ dW_t \\ 
z^5_t\circ dW_t \\
-z^8_t\circ dW_t \\
z^7_t\circ dW_t
\end{bmatrix},\quad\cdots, \quad d\bar{z}_{7\:t}=
\begin{bmatrix}
-z^8_t\circ dW_t \\
-z^7_t\circ dW_t\\
-z^6_t\circ dW_t\\
-z^5_t\circ dW_t\\
z^4_t\circ dW_t\\
z^3_t\circ dW_t\\
z^2_t\circ dW_t\\
z^1_t\circ dW_t
\end{bmatrix}.
\label{sde17}
\end{eqnarray}
Since $U_1, U_2, \cdots, \text{ and }U_7$ are Killing vector fields, the flows generated by equation \eqref{sde16} are isometric. The isometric stochastic flows $\bar{z}_{\mu \:t}$ will be called {\em frame isometric stochastic flows}. Due to equation \eqref{sde17}, the process $\bar{z}_{\mu \:t}$ is a semimartingale for all $\mu =1, \cdots, 7$.

An arbitrary vector field $A$ on $S^7_s$ can be written as a linear combination of the frame $\{U_1, \cdots, U_7\}$, namely $A=A^{\mu} U_{\mu }$ (Einstein convention of summation). Therefore, the isometric stochastic flow $z_t$ generated by $A$ is given by
\begin{equation}
dz_t=A(z_t)\circ dW_t.
\label{sde18}
\end{equation}
From equation \eqref{sde16}, the last expression can be written as
\begin{equation}
dz_t=A(z_t)\circ dW_t=A^{\mu }(z_t)U_{\mu}(z_t)\circ dW_t.
\label{sde19}
\end{equation}
The above vector field $A$ is a Killing vector field if $\mathcal{L}_{A^\mu U_\mu}g=0$, i.e.,
\begin{eqnarray}
\big(\mathcal{L}_{A^\mu U_\mu}g\big)_{ij}&=& \big(A^\mu \mathcal{L}_{U^\mu}\: g\big)_{ij}+\big(\partial_i \:A^\mu\big) U_\mu^k \:g_{kj}+\big(\partial_j A^\mu\big) U_\mu^k\:g_{ik}\nonumber\\
&=&\big(A^\mu \mathcal{L}_{U^\mu}\: g\big)_{ij}+U_\mu^j\;\partial_i \:A^\mu+U_\mu^i\;\partial_j A^\mu\nonumber\\
&=&0.
\end{eqnarray}
Since $U^\mu (\mu=1, 2, \cdots, 7)$ are Killing vector fields, then $A$ is also Killing vector field if
\begin{equation}
U_\mu^j\;\partial_i \:A^\mu=- U_\mu^i\;\partial_j A^\mu,
\label{Killing vector field 21}
\end{equation}
for all $i, j=1, 2, \cdots, 8$. Therefore, the stochastic flow generated by equation \eqref{sde19} is isometric if the vector fields $A$ satisfies equation \eqref{Killing vector field 21}.

\subsubsection{\textbf{Fokker-Planck equations on $S^7_s$ associated to frame isometric stochastic flow and their entropy rate}}

Equation \eqref{sde16} can be written in matrix notation as follows:
\begin{eqnarray}
\label{Stra}
\begin{bmatrix}
d\bar{z}_{{\mu }\:t}^1\\
\\
d\bar{z}_{{\mu }\:t}^2\\
\\
\vdots\\
\\
d\bar{z}_{{\mu }\:t}^8
\end{bmatrix}
&=&
\begin{bmatrix}
U_\mu^1&0&0&0&0&0&0&0\\
0&U_\mu^2&0&0&0&0&0&0\\
0&0&U_\mu^3&0&0&0&0&0\\
0&0&0&U_\mu^4&0&0&0&0\\
0&0&0&0&U_\mu^5&0&0&0\\
0&0&0&0&0&U_\mu^6&0&0\\
0&0&0&0&0&0&U_\mu^7&0\\
0&0&0&0&0&0&0&U_\mu^8\\
\end{bmatrix}
\circ 
\begin{bmatrix}
dW_t\\
\\
dW_t\\
\\
\vdots\\
\\
dW_t\\
\end{bmatrix}\nonumber\\
d\bar{z}_{\mu\:t}^{\;i}&=&\sum_{k=1}^{8}\delta_{i\:k}\:U_\mu^i (\bar{z}_t)\circ dW_t,
\label{sde21}
\end{eqnarray}
where $U_\mu^1, U_\mu^2, \cdots, U_\mu^8$ are the components of $U_\mu$. The Stratonovich integral of stochastic differential equation \eqref{sde21} is given by
\begin{equation}
\int_{t_0}^{t}\sum_{k=1}^{8}\delta_{i\:k}\:U_\mu^i(\bar{z}_s)\circ dW_s=\int_{t_0}^{t} \sum_{k=1}^{8}\delta_{i\:k}\:U_\mu^i(\bar{z}_s) dW_s+\frac{1}{2}\int_{t_0}^{t} h^i(\bar{z}_s) ds.
\label{drifsde22}
\end{equation}
Note that as long as the diffusion function $\delta_{i\:k}\:U_\mu^i(\bar{z}_t)$ is only dependent on the random field then It\^{o} and Stratonovich interpretations of the stochastic differential equation are the same. The modified drift function (It\^{o}-Stratonovich correction) in equation \eqref{drifsde22} is given by (e.g., \cite{Chirikjian}) 
\begin{eqnarray}
h^i(\bar{z}_t)&=&\sum_{k=1}^{8}\big(\nabla\cdot [\delta_{i\:k}\:U_\mu^i(\bar{z}_t)\; \delta_{k\:j}\:U_\mu^j (\bar{z}_t)]-\delta_{i\:k}\:U_\mu^i (\bar{z}_t) \nabla\cdot \delta_{k\:j}\:U_\mu^j (\bar{z}_t)\big)\nonumber\\
&=&\sum_{k=1}^{8}\delta_{k\:j}\:U_\mu^j(\bar{z}_t)\frac{\partial \delta_{i\:k}\:U_\mu^i (\bar{z}_t)}{\partial \bar{z}_t^k}.
\label{drif23}
\end{eqnarray}
In fact, the It\^{o} counterpart of the Stratonovich stochastic differential equation is a useful artifice which, for example, allows access to the Fokker-Planck equation. Similarly, we must use the corresponding It\^{o} stochastic differential equation to determine the appropriate coefficients of the Fokker-Planck equation for a diffusion process arising from the solution of a Stratonovich equation.

Therefore, if a physical system, on the one hand, is defined by the Stratonovich stochastic differential equations \eqref{sde21} then the same process can also be described by the It\^{o} equation:
\begin{equation} \label{sde24}
d\bar{z}_{{\mu }\:t}^{\:i}=\frac{1}{2}\sum_{k=1}^{8}\delta_{k\:j}\:U_\mu^j (\bar{z}_t)\frac{\partial \delta_{i\:k}\:U_\mu^i (\bar{z}_t)}{\partial \bar{z}_t^k} dt+\sum_{k=1}^{8}\delta_{i\:k}\:U_\mu^i (\bar{z}_t) dW_t,
\end{equation}
with the solution is given by
\begin{equation}
\bar{z}_{\mu\:t}^{\:i}=\bar{z}_{\mu\: t_0}^{\:i}+\frac{1}{2} \int_{t_0}^{t}\sum_{k=1}^{8}\delta_{k\:j}\:U_\mu^j(\bar{z}_s)\frac{\partial \delta_{i\:k}\:U_\mu^i (\bar{z}_s)}{\partial \bar{z}_t^k}ds+\int_{t_0}^{t}\sum_{k=1}^{8}\delta_{i\:k}\:U_\mu^i (\bar{z}_s) dW_s.
\label{Itosde25}
\end{equation}
It\^{o}-Stratonovich stochastic differential equation and Fokker-Planck equations can be formulated to describe the dynamical system in any coordinate of a Riemannian manifold in a way that is very similar to the case of $\mathbb{R}^d$. Therefore, the derivation of the Fokker-Planck equation governing the time evolution of conditional probability density function on $S^7_s$ proceeds in an analogous way to the derivation of such equation in $\mathbb{R}^8$. In general, the volume of $S^7_s$ will be an $7$-dimensional measure.

Let us assume that $f(\bar{z}_{\mu\:s}, s)$ is sufficiently differentiable function, with $f(\bar{z}_{\mu\:s}, s)=0$ for $s\notin (t, t_0)$. In this note, all expectations are conditional expectations with respect to $\bar{z}_{\mu\:t_0}$; so that 
\begin{equation}
\mathbb{E}(f(\bar{z}_{\mu\:t}, t) |\bar{z}_{\mu\:t_0}, t_0)=\int_{\mathbb{R}^8} f(\bar{z}_{\mu\:t}, t) p(\bar{z}_{\mu\:t}, t|\bar{z}_{\mu\:t_0}, t_{0}) d\bar{z}_{\mu\:t},
\label{expectation26}
\end{equation}
where $\mu=1,2,\cdots,7$. If $(\phi_1, \phi_2, \phi_3, \phi_4, \phi_5, \phi_6, \phi_7)$ is an $7$-spherical coordinates system, then the Jacobian determinant associate to the change of coordinate system from Cartesian coordinates-$\mathbb{R}^8$ is given by
\begin{eqnarray}
d_{\mathbb{R}^8}V&=&\bigg|\det \bigg[\frac{\partial z_{\mu}}{\partial r}, \frac{\partial z_{\mu}}{\partial \phi_1}, \cdots, \frac{\partial z_{\mu}}{\partial \phi_{7}}\bigg]\bigg| dr \;d\phi_1 \;d\phi_2\cdots \;d\phi_{7}\nonumber\\
&=&d_{S^7_s}V(\phi_1, \phi_2, \cdots, \phi_7) r^{7}dr,
\label{volume27}
\end{eqnarray}
such that equation \eqref{expectation26} can be rewritten as
\begin{eqnarray}
&&\mathbb{E}(f(\phi_1, \phi_2, \cdots, \phi_7; t) |\phi_1, \phi_2, \cdots, \phi_7; t_0)\nonumber\\
&&=\int_{S^7_s} f(\phi_1, \cdots, \phi_7; t) p(\phi_1, \cdots, \phi_7; t|\phi'_1, \cdots, \phi'_7; t_{0}) \;d_{S^7_s}V(\phi_1, \cdots, \phi_7),
\label{expectation27}
\end{eqnarray}
where $\phi'_1=\phi_1(t-dt)$ and $d_{S^7_s}V(\phi_1, \cdots, \phi_7)$ is the volume element for the sphere $S^7_s$. The volume element of $S^7_s$, is given by
\begin{eqnarray}
d_{S^7_s}V(\phi_1, \cdots, \phi_7)&=&\sin^6(\phi_1)\:\sin^5(\phi_2)\:\sin^4(\phi_3)\:\sin^3(\phi_4)\:\sin^2(\phi_5)\:\sin(\phi_6)\:d\phi_1\:d\phi_2\:d\phi_3\:d\phi_4\:d\phi_5\:d\phi_6\:d\phi_7\nonumber\\
&=&\prod_{p=1}^6 \sin^{7-p}(\phi_p)\prod_{q=1}^7\: d(\phi_q),
\label{Volume29}
\end{eqnarray}
where the angles $\{\phi_{1},\phi_{2},\dots, \phi_{6}\}$ range over $[0,\pi]$ radians and angle $\{\phi_{7}\}$ ranges over $[0, 2\pi]$ radians.

Differential volume elements described in the two coordinate systems are related by the expression $d(z_{\mu})=|J(\phi_q)| d(\phi_q)$, $q=1, 2, \cdots, 7$. In this paper, the matrix $G(\phi_q)=J^T(\phi_q) J(\phi_q)$ (called the metric tensor) contains all of the information needed to measure distances, areas, volumes, etc. Since $|J(\phi_q)|=|G(\phi_q)|^{\frac{1}{2}}$, $q=1, 2, \cdots, 7$, the volume element in spherical coordinates can be expressed as $d(z_{\mu})=|G(\phi_q)|^{\frac{1}{2}} d(\phi_q)=\prod_{p=1}^6 \sin^{7-p}(\phi_p)$ $\prod_{q=1}^7\: d(\phi_q)$.

The derivation of the Fokker-Planck equation in Stratonovich version is not as simple as for the It\^{o} stochastic differential equation, since the expected value of a Strantonovich integral is, in general, non-zero, but one can show that the Fokker-Planck equation for Stratonovich version \eqref{sde21} is
\begin{eqnarray}
&&\frac{\partial p(\phi_1, \cdots, \phi_7; t|\phi'_1, \cdots, \phi'_7; t_{0})}{\partial t}\; \prod_{p=1}^6 \sin^{7-p}(\phi_p)\nonumber\\
&=&-\frac{1}{2}\sum_{i=1}^{7}\frac{\partial}{\partial \phi_i}\bigg[\sum_{j,\:k=1}^{7}\delta_{kj}\:U_\mu^j (\phi_1, \cdots, \phi_7)\frac{\partial \delta_{ik}\:U_\mu^i (\phi_1, \cdots, \phi_7)}{\partial \phi_k} p(\phi_1, \cdots, \phi_7; t|\phi'_1, \cdots, \phi'_7; t_{0})\prod_{p=1}^6 \sin^{7-p}(\phi_p)\bigg]\nonumber\\
&&+\frac{1}{2}\sum_{i,\: j=1}^7 \frac{\partial^2}{\partial\phi_i\:\partial \phi_j}\bigg[\sum_{k=1}^{7}\delta_{i\:k}\:U_\mu^i (\phi_1\cdots, \phi_7)\; \delta_{k\:j}\:U_\mu^j (\phi_1\cdots, \phi_7) \times p(\phi_1, \cdots, \phi_7; t|\phi'_1, \cdots, \phi'_7; t_{0})\nonumber\\
&&\qquad\prod_{p=1}^6 \sin^{7-p}(\phi_p)\bigg].
\label{Fokker30}
\end{eqnarray}

Entropy is a concept which has played a central role in several areas such as statistical mechanics and information theory. Entropy is a measure of unpredictability associated with random phenomena. The entropy of the conditional transition probability density function $p(\phi_1, \cdots, \phi_7; t|\phi'_1, \cdots, \phi'_7; t_{0})$ describing the distribution of states of a random field $z$ on $S^7_s$ is defined by the integral \cite{Chirikjian}
\begin{eqnarray}
&&S(p(\phi_1, \cdots, \phi_7; t|\phi'_1, \cdots, \phi'_7; t_{0}))\nonumber\\
&=&-\int_{S^7_s}\bigg[p(\phi_1, \cdots, \phi_7; t|\phi'_1, \cdots, \phi'_7; t_{0})\log p(\phi_1, \cdots, \phi_7; t|\phi'_1, \cdots, \phi'_7; t_{0})\bigg]\prod_{p=1}^6 \sin^{7-p}(\phi_p)\prod_{q=1}^7\: d(\phi_q).
\label{entropy31}
\end{eqnarray}
Quantities of the form $S(p(\phi_1, \cdots, \phi_7; t|\phi'_1, \cdots, \phi'_7; t_{0}))$ is information-theoretic entropy playing a central role in information theory as measures of information, choice and uncertainty.

Differentiating \eqref{entropy31} with respect to time gives
\begin{eqnarray}
\frac{dS}{dt}&=&-\int_{S^7_s}\bigg[\frac{\partial p(\phi_1, \cdots, \phi_7; t|\phi'_1, \cdots, \phi'_7; t_{0})}{\partial t} \log p(\phi_1, \cdots, \phi_7; t|\phi'_1, \cdots, \phi'_7; t_{0})\nonumber\\
&&+\frac{\partial p(\phi_1, \cdots, \phi_7; t|\phi'_1, \cdots, \phi'_7; t_{0})}{\partial t}\bigg]\;\prod_{p=1}^6 \sin^{7-p}(\phi_p)\prod_{q=1}^7\: d(\phi_q).
\label{entropy32}
\end{eqnarray}
From integration by parts, the second integral in equation \eqref{entropy32} vanishes due to the fact that the conditional transition probability density is preserved by the Fokker-Planck equation. The remaining term can be written as
\begin{eqnarray}
\frac{dS}{dt}&=&-\int_{S^7_s}\bigg[\frac{\partial p(\phi_1, \cdots, \phi_7; t|\phi'_1, \cdots, \phi'_7; t_{0})}{\partial t} \log p(\phi_1, \cdots, \phi_7; t|\phi'_1, \cdots, \phi'_7; t_{0}) \bigg] \prod_{p=1}^6 \sin^{7-p}(\phi_p)\prod_{q=1}^7\: d(\phi_q)\nonumber\\
&=&\int_{S^7_s}\Bigg\{-\frac{1}{2}\sum_{i,\: j=1}^7 \frac{\partial^2}{\partial\phi_i\:\partial \phi_j}\bigg[\prod_{p=1}^6 \sin^{7-p}(\phi_p)\sum_{k=1}^{7} \delta_{i\:k}\:U_\mu^i (\phi_1\cdots, \phi_7)\;  \delta_{k\:j}\:U_\mu^j (\phi_1\cdots, \phi_7)\nonumber\\
&&\qquad\;p(\phi_1, \cdots, \phi_7; t|\phi'_1, \cdots, \phi'_7; t_{0})\bigg]\Bigg\}\log p(\phi_1, \cdots, \phi_7; t|\phi'_1, \cdots, \phi'_7; t_{0}) \prod_{q=1}^7\: d(\phi_q).
\label{entropy33}
\end{eqnarray}
Integrating by parts and ignoring the boundary terms gives the rate of the change of entropy as follows
\begin{eqnarray}
\frac{dS}{dt}&=&-\int_{S^7_s}\Bigg\{\frac{1}{2}\sum_{i,\: j=1}^7\bigg[-\frac{1}{p(\phi_1, \cdots, \phi_7; t|\phi'_1, \cdots, \phi'_7; t_{0})}\sum_{k=1}^{7} \delta_{i\:k}\:U_\mu^i (\phi_1\cdots, \phi_7)\: \delta_{k\:j}\:U_\mu^j (\phi_1\cdots, \phi_7)\nonumber\\
&&\quad\bigg(\frac{\partial p(\phi_1, \cdots, \phi_7; t|\phi'_1, \cdots, \phi'_7; t_{0})}{\partial \phi_i}\;\frac{\partial p(\phi_1, \cdots, \phi_7; t|\phi'_1, \cdots, \phi'_7; t_{0})}{\partial \phi_j}\bigg)\nonumber\\
&&\quad+\frac{\partial^2 p(\phi_1, \cdots, \phi_7; t|\phi'_1, \cdots, \phi'_7; t_{0})}{\partial\phi_i\:\partial \phi_j}\nonumber\\
&&\;\;\sum_{k=1}^{7}\delta_{i\:k}\:U_\mu^i (\phi_1\cdots, \phi_7)\; \delta_{k\:j}\:U_\mu^j (\phi_1\cdots, \phi_7)\bigg]\Bigg\}\prod_{p=1}^6 \sin^{7-p}(\phi_p)\prod_{q=1}^7\: d(\phi_q).
\label{entropy34}
\end{eqnarray}

	\subsubsection{\textbf{Fokker-Planck equation generated by an arbitrary vector field and its entropy rate}}
    Equation \eqref{sde19} can be written in matrix notation as follows:
\begin{eqnarray}
\begin{bmatrix}
dz_t^1\\
\\
dz_t^2\\
\\
\vdots\\
\\
dz_t^8
\end{bmatrix}&=&\begin{bmatrix}
U^1_1\;A^1\qquad&U^1_2\;A^2\qquad&\;\cdots\;\qquad&U^1_7\;A^7\\
\\
U^2_1\;A^1\qquad&U^2_2\;A^2\qquad&\;\cdots\;\qquad&U^2_7\;A^7\\
\\
\vdots\qquad&\vdots\qquad&\vdots\qquad&\vdots\\
\\
U^8_1\;A^1\qquad&U^8_2\;A^2\qquad&\;\cdots\;\qquad&U^8_7\;A^7
\end{bmatrix}\circ\begin{bmatrix}
dW_t\\
\\
dW_t\\
\\
\vdots\\
\\
dW_t\\
\end{bmatrix}\nonumber\\
dz_t^i&=&\sum_{\mu=1}^{7}U_{\mu}^i(\bar{z}_t)\:A^\mu(z_t)\circ dW_t,
\label{sde35}
\end{eqnarray}
where $U_\mu^1 A^\mu, U_\mu^2 A^\mu, \cdots, U_\mu^8 A^\mu$ (without summation in index $\mu$) are the components of a matrix $UA$, $(UA)_{i\mu}=U^i_\mu A^\mu$. The Stratonovich integral of stochastic differential equation \eqref{sde35} is given by
\begin{equation}
\int_{t_0}^{t}\sum_{\mu=1}^{7}(UA)_{i\:\mu}(z_s)\circ dW_s=\int_{t_0}^{t}\sum_{\mu=1}^{7}(UA)_{i\:\mu}(z_s) dW_s+\frac{1}{2}\int_{t_0}^{t} h^i(z_s) ds.
\label{sde36}
\end{equation}
The modified drift function (It\^{o}-Stratonovich correction) in equation \eqref{sde36} is given by
\begin{equation}
h^i(z_t)=\sum_{\mu=1}^{7}(UA)_{\mu\: j}(z_t)\frac{\partial (UA)_{i\:\mu}(z_t)}{\partial z_t^\mu},
\label{driftsde37}
\end{equation}
so that
\begin{equation} 
dz_t^i=\frac{1}{2}\sum_{\mu=1}^{7}(UA)_{\mu\: j}(z_t)\frac{\partial (UA)_{i\:\mu}(z_t)}{\partial z_t^\mu} dt+\sum_{\mu=1}^{7}(UA)_{i\:\mu}(z_t)dW_t.
\label{drift38}
\end{equation}
By using the volume element of $S^7_s$ given by equation \eqref{Volume29} we obtain the Fokker-Planck equation for Stratonovich equation \eqref{sde35} as follows
\begin{eqnarray}
&&\prod_{p=1}^6 \sin^{7-p}(\phi_p)\frac{\partial p(\phi_1, \cdots, \phi_7; t|\phi'_1, \cdots, \phi'_7; t_{0})}{\partial t}\nonumber\\
&=&-\frac{1}{2}\sum_{i=1}^{7}\frac{\partial}{\partial \phi_i}\bigg[\sum_{j=1}^{7}\sum_{\mu=1}^{7}(UA)_{\mu\: j}(\phi_1, \cdots, \phi_7)\frac{\partial (UA)_{i\:\mu}(\phi_1, \cdots, \phi_7)}{\partial \phi_\mu}\;p(\phi_1, \cdots, \phi_7; t|\phi'_1, \cdots, \phi'_7; t_{0})\nonumber\\
&&\qquad \prod_{p=1}^6 \sin^{7-p}(\phi_p)\bigg]+\frac{1}{2}\sum_{i,\: j=1}^7 \frac{\partial^2}{\partial\phi_i\:\partial \phi_j}\bigg[\sum_{\mu=1}^{7}(UA)_{i\:\mu}(\phi_1, \cdots, \phi_7)\;(UA)_{\mu\: j}(\phi_1, \cdots, \phi_7)\nonumber\\
&&\qquad \;p(\phi_1, \cdots, \phi_7; t|\phi'_1, \cdots, \phi'_7; t_{0})\prod_{p=1}^6 \sin^{7-p}(\phi_p)\bigg].
\label{FokkerPlanck39}
\end{eqnarray}

The entropy rate of the process on $S^7_s$ is defined by the integral related to Fokker-Planck equation \eqref{FokkerPlanck39} is given by
\begin{eqnarray}
\frac{dS}{dt}&=&-\int_{S^7_s}\bigg( \frac{\partial p(\phi_1, \cdots, \phi_7; t|\phi'_1, \cdots, \phi'_7; t_{0})}{\partial t}\;\log p(\phi_1, \cdots, \phi_7; t|\phi'_1, \cdots, \phi'_7; t_{0}) \bigg)\prod_{p=1}^6 \sin^{7-p}(\phi_p)\;\prod_{q=1}^7\: d(\phi_q)\nonumber\\
&=&\int_{S^7_s}\Bigg\{-\frac{1}{2}\sum_{i,\: j=1}^7 \frac{\partial^2}{\partial\phi_i\:\partial \phi_j}\bigg[\prod_{p=1}^6 \sin^{7-p}(\phi_p)\sum_{\mu=1}^{7}(UA)_{i\:\mu}(\phi_1, \cdots, \phi_7)\; (UA)_{\mu\: j} (\phi_1, \cdots, \phi_7) \nonumber\\
&&\qquad p(\phi_1, \cdots, \phi_7; t|\phi'_1, \cdots, \phi'_7; t_{0})\bigg]\Bigg\}\log p(\phi_1, \cdots, \phi_7; t|\phi'_1, \cdots, \phi'_7; t_{0}) \prod_{q=1}^7\: d(\phi_q).
\label{entropy40}
\end{eqnarray}

Integrating by parts and ignoring the boundary terms gives the rate of the change of entropy as follows
\begin{eqnarray}
\frac{dS}{dt}&=&-\int_{S^7_s}\Bigg\{\frac{1}{2}\sum_{i,\: j=1}^7\bigg[-\frac{1}{p(\phi_1, \cdots, \phi_7; t|\phi'_1, \cdots, \phi'_7; t_{0})}\sum_{\mu=1}^{7}(UA)_{i\:\mu}(\phi_1, \cdots, \phi_7)\:(UA)_{\mu\: j}(\phi_1, \cdots, \phi_7)\nonumber\\
&&\qquad\bigg(\frac{\partial p(\phi_1, \cdots, \phi_7; t|\phi'_1, \cdots, \phi'_7; t_{0})}{\partial \phi_i}\frac{\partial p(\phi_1, \cdots, \phi_7; t|\phi'_1, \cdots, \phi'_7; t_{0})}{\partial \phi_j}\bigg)\nonumber\\
&&+\frac{\partial^2 p(\phi_1, \cdots, \phi_7; t|\phi'_1, \cdots, \phi'_7; t_{0})}{\partial\phi_i\:\partial \phi_j}\sum_{\mu=1}^{7}(UA)_{i\:\mu}(\phi_1, \cdots, \phi_7)(UA)_{\mu\: j}(\phi_1, \cdots, \phi_7)\bigg]\Bigg\}\nonumber\\
&&\prod_{p=1}^6 \sin^{7-p}(\phi_p)\;\prod_{q=1}^7\: d(\phi_q).
\label{entropy41}
\end{eqnarray}

\subsection{Stochastic processes on $\Sigma^7_{GM}$}

Topologically speaking, two objects are said to be equivalent if there is a homeomorphism from one onto another. A homeomorphism is a one-to-one onto a map that is continuous, and whose inverse is also continuous. It is a topological isomorphism. Note that $S^7_s$ and $\Sigma^7_{GM}$ are homeomorphic but not diffeomorphic. Therefore, there must exist some homeomorphisms $h: S^7_s\rightarrow \Sigma^7_{GM}$ from $S^7_s$ onto $\Sigma^7_{GM}$. We will construct one of such homeomorphisms and apply it to study the induced isometric stochastic flows on $\Sigma^7_{GM}$. It means that we will consider some isometric stochastic flows on the $S^7_s$ however, in another class of differential structure which is not equivalent to the standard one.

The situation can be described as follows. Both manifolds $S^7_s$ and $\Sigma ^7_{GM}$ can be regarded as two topological spaces embedded in $\mathbb{R}^8$. The topological space $S^7_s$ is the seven-dimensional sphere $S^7$ defined by the standard or natural metric in $\mathbb{R}^8$. The identification \eqref{identification10} can be regarded as the inclusion of $S^7$ into $\mathbb{R}^8$. The standard sphere $S^7_s$ is constructed from the seven-sphere $S^7$ equipped with the differential structure which is inherited by $S^7$ from the standard differential structure in $\mathbb{R}^8$. Whereas the manifold $\Sigma ^7_{GM}$ is constructed from the topological space $\mathrm{Sp}(2, \mathbb{H})/\star $ also embedded in $\mathbb{R}^8$ equipped with the differential structure which is inherited from the standard differential structure in $\mathbb{R}^8$. In this description, it is clear that the topological space $\mathrm{Sp}(2, \mathbb{H})/\star $ as a subset of $\mathbb{R}^8$ cannot be the seven-sphere $S^7$.

Now consider again both actions of $S^3\cong \mathrm{Sp}(1, \mathbb{H})$ on $\mathrm{Sp}(2, \mathbb{H})$. In fact, the fibers of both actions in $\mathrm{Sp}(2, \mathbb{H})$ are coincide for elements in $\mathrm{Sp}(2, \mathbb{H})$ of the form (see \cite{Gromoll})
\begin{equation}
\begin{pmatrix}
\alpha&\beta\\
-\beta &\alpha
\end{pmatrix} \qquad\qquad \alpha, \beta\in\mathbb{R}.
\label{fiber42}
\end{equation}
Let $R$ be a subset of $\mathrm{Sp}(2, \mathbb{H})$ of the form given by equation \eqref{fiber42}. For the standard action, the fiber is determined by \cite{Speranca}
\begin{equation}
q\bullet \begin{pmatrix}
\alpha&\beta\\
-\beta &\alpha
\end{pmatrix}=\begin{pmatrix}
\alpha&\beta\bar{q}\\
-\beta &\alpha\bar{q}
\end{pmatrix}=
\begin{pmatrix}
\alpha&\bar{q}\beta\\
-\beta &\bar{q}\alpha
\end{pmatrix},
\label{dott43}
\end{equation}
for every $q\in \mathrm{Sp}(1, \mathbb{H})$. Whereas for the Gromoll-Meyer exotic action, the fiber is obtained as follows
\begin{equation}
q\star\begin{pmatrix}
\alpha&\beta\\
-\beta &\alpha
\end{pmatrix}=
\begin{pmatrix}
q\alpha\bar{q}&q\beta\\
-q\beta\bar{q} &q\alpha
\end{pmatrix}=\begin{pmatrix}
\alpha&q\beta\\
-\beta &q\alpha
\end{pmatrix},
\label{star44}
\end{equation}
for every $q\in \mathrm{Sp}(1, \mathbb{H})$. Therefore, we have
\begin{equation}
\begin{bmatrix}
\begin{pmatrix}
\alpha&\beta\\
-\beta &\alpha
\end{pmatrix}
\end{bmatrix}^\bullet
=
\begin{bmatrix}
\begin{pmatrix}
\alpha&\beta\\
-\beta &\alpha
\end{pmatrix}
\end{bmatrix}^\star,
\label{equivalent45}
\end{equation}
for every $(\alpha, \beta)\in\mathbb{R}^2$, where $[ \quad ]^\bullet$ and $[ \quad ]^\star$ denote fibers of the standard action and the Gromoll-Meyer exotic action, respectively.

Furthermore, from the identification equation \eqref{identification10}, the class
\begin{equation}
\begin{bmatrix}
\begin{pmatrix}
\alpha&\beta\\
-\beta &\alpha
\end{pmatrix}
\end{bmatrix}^\bullet,
\label{circle46}
\end{equation}
is identified with the point $(\beta, \alpha, 0,0,0,0,0,0)\in S^7_s \subset \mathbb{R}^8$. Since $\beta ^2+\alpha ^2=1$ it is clear that the set $V_\bullet$ of all fibers of $\bullet$-fibration containing all elements of $\mathrm{Sp}(2, \mathbb{H})$ of the form given by equation \eqref{circle46} is a circle on $S^7_s$. It is obtained by intersecting $S^7_s$ by the $(z^1z^2)$-plane. The set $V_\bullet$ is therefore closed in $S^7_s$. From the identification \eqref{equivalent45}, we see that the set $V_\star $ of of all fibers of $\star$-fibration containing all elements of $\mathrm{Sp}(2, \mathbb{H})$ of the form given by equation \eqref{equivalent45} is the same circle on $S^7_s$. Both topological space as subset of $\mathbb{R}^8$ at least have a unit circle on the $(z^1z^2)$-plane of $\mathbb{R}^8$ as a part of the intersection of both topological space.

Therefore we have the identity map $\text{Id }: V_\bullet\rightarrow V_\star$ from the closed subset $V_\bullet$ onto the closed subset $V_\star$ of all fibers of $\star$-fibration containing elements of $\mathrm{Sp}(2, \mathbb{H})$ of the form given by equation \eqref{fiber42}. Since $V_\bullet$ is closed then from the extension theorem of a mapping on a closed set there exist (not uniquely) an extensions of the identity map $\text{Id }: V_\bullet\rightarrow V_\star$ to the whole space $S^7_s$. Here, an extension of $\text{Id}$ to $S^7_s$ which is a homeomorphism will be denoted by $h$.

Now consider the set $S^1_{23}$ of all classes of $\bullet $-fibration containing elements of $\mathrm{Sp}(2, \mathbb{H})$ of the form given by
\begin{equation}
\begin{bmatrix}
\begin{pmatrix}
\alpha&\beta \mathrm{i}\\
\beta \mathrm{i} &\alpha
\end{pmatrix}
\end{bmatrix}^\bullet.
\label{circle47}
\end{equation}
It is clear that the set is a circle on $S^7_s$ whose points are represented by $(0,\beta, \alpha, 0, 0, 0, 0, 0)\in S^7_s\subset \mathbb{R}^8$ after the identification given in equation \eqref{identification10}. The circle is obtained by intersecting $S^7_s$ by the $(z^2z^3)$-plane. Since $\beta \mathrm{i}$ does not commute with any quaternion $q$, the fiber of the $\star $-fibration in $\mathrm{Sp}(2,\mathbb{H})$ which is identified with the above fiber of $\bullet $-fibration by the homeomorphism $h$ is {\em not} of the form given by 
\begin{equation}
\begin{bmatrix}
\begin{pmatrix}
\alpha&\beta \mathrm{i}\\
\beta \mathrm{i} &\alpha
\end{pmatrix}
\end{bmatrix}^\star.
\label{identification48}
\end{equation}
Therefore, keeping in mind the above view in which both topological spaces are regarded as subsets of $\mathbb{R}^8$, the image of the circle $S^1_{23}$ on $\Sigma ^7_{GM}$ under the homeomorphism $h$ may not be the circle itself. Surely, it is a closed curve in $\Sigma ^7_{GM}$.

Similarly, we obtain the other such circles on $S^7_s$. There are twenty eight such circles including $ V_\bullet: =S^1_{12}$. For $i,j\in \{1,\cdots, 8\}$ with $i<j$, the circle $S^1_{ij}$ corresponds to an integral curve of the vector field $U_{ij}$ appearing in the expressions of vector fields $U's$ above. The images of each circle under the homeomorphism $h$ are closed curves on $\Sigma^7_{GM}$ which are in general not circles on $S^7_s$. The set of twenty-eight closed curves on $\Sigma ^7_{GM}\subset \mathbb{R}^8$ related to the twenty-eight circles on $S^7_s$ by the homeomorphism $h$ frames the manifold $\Sigma^7_{GM}$.

Whenever it is needed, the topological space $\mathrm{Sp}(2, \mathbb{H})/\star $ can be reshaped by using a differentiable deformation so that every line emerging from the origin of $\mathbb{R}^8$ intersects the deformed topological space only in one point, while the circle $S^1_{12}$ in $\Sigma^7_{GM}$ does not change. Then, the homeomorphism $h$ can be defined for instance by
\begin{equation}
h(z^1,\cdots ,z^8)=D^{-1}[\beta (z^1,\cdots ,z^8)(z^1,\cdots ,z^8)],
\end{equation}
for every $(z^1,\cdots ,z^8)\in S^7_s$ where $D:\mathbb{R}^8\rightarrow \mathbb{R}^8$ is the above mentioned differentiable deformation with $D(\gamma ^1,\cdots ,\gamma ^8)\neq (0,\cdots ,0)$ for all $(\gamma ^1,\cdots ,\gamma ^8)\in \Sigma ^7_{GM}$ and $\beta : S^7_s\rightarrow \mathbb{R}$ is a suitable real positive function on $S^7_s$. The function $\beta $ is at least continuous and carries therefore the exotism of $\Sigma ^7_{GM}$. If
\begin{equation}
(\zeta ^1,\cdots \zeta ^8):=\beta (z^1,\cdots ,z^8)(z^1,\cdots ,z^8),
\end{equation}
then
\begin{equation}
h(z^1,\cdots ,z^8)=D^{-1}(\zeta ^1,\cdots \zeta ^8).
\end{equation}
Since $(z^1)^2+\cdots +(z^8)^2=1$, then
\begin{equation}
\beta (z^1,\cdots ,z^8)=[(\zeta ^1)^2+\cdots +(\zeta ^8)^2]^{1/2},
\end{equation}
and
\begin{equation}
h^{-1}(\gamma ^1,\cdots ,\gamma ^8)=|D(\gamma ^1,\cdots \gamma ^8)|^{-1}D(\gamma ^1,\cdots \gamma ^8),
\end{equation}
for every $(\gamma ^1,\cdots \gamma ^8)\in \Sigma ^7_{GM}$, where $|D(\gamma ^1,\cdots \gamma ^8)|$ is given by
\begin{equation}
|D(\gamma ^1,\cdots \gamma ^8)|=[(D^1)^2+\cdots +(D ^8)^2]^{1/2}=[(\zeta ^1)^2+\cdots +(\zeta ^8)^2]^{1/2}.
\end{equation}

The homeomorphism $ h: S^7_s\rightarrow \Sigma^7_{GM} $ which has been constructed in the previous part identifies the existing classes in $S^7_s$ and those in $ \Sigma^7_{GM} $. A point $(b,d)=(z^1, z^2, \cdots, z^8)\in S^7_s$ represents the point $h(b,d)$ in $\Sigma^7_{GM} $. It is clear that if the stochastic flows under consideration are a stochastic flow of homeomorphism, then, the stochastic flows on $\Sigma^7_{GM}$ are equivalent to the stochastic flows on $S^7_s$. However, it may not be so in the case of a stochastic flow of diffeomorphism.

In this work, we assume that a stochastic flow of diffeomorphism $g_{s, t}(\omega)$ on $S^7_s$ is given (see Subsection 4.1.2). Then we can construct a stochastic flow on $\Sigma^7_{GM}$ and study the properties of the flow. Let the stochastic flow of diffeomorphism $g_{s,t}(\omega )$ be the solution of the above Stratonovich stochastic differential equation \eqref{sde13}. Using the homeomorphism $h$, then 
\begin{equation}\label{inducedflow}
h_\ast g_{s,t}(\omega ):=h\circ g_{s,t}(\omega )\circ h^{-1},
\end{equation}
is a stochastic flow on $\Sigma^7_{GM}$. The stochastic flow $h_\ast g_{s,t}(\omega )$ can be regarded as the same stochastic flow $g_{s,t}(\omega )$ on the seven sphere $S^7$, but viewed in exotic (Gromoll-Meyer) differential structure. Clearly, all topological properties appearing in the stochastic flow of diffeomorphism $g_{s,t}(\omega )$ on $S^7_s$ are observed also in the stochastic flow $h_\ast g_{s,t}(\omega )$ on $\Sigma^7_{GM}$. If the map $h$ was a diffeomorphism, the flow $h_\ast g_{s,t}(\omega )$ is clearly the unique solution of the following stochastic differential equation
\begin{equation}
d(h\circ z_t\circ h^{-1})=h_\ast V_{\alpha }(h\circ z_t\circ h^{-1})\circ dW_{t}^{\alpha }+h_\ast V_0(h\circ z_t\circ h^{-1})dt,
\label{sdeh}
\end{equation}
and therefore the flow $h_\ast g_{s,t}(\omega )$ has the same regularities as those of the flow $g_{s,t}(\omega )$. However, the vector fields $h_\ast V_0$ and $h_\ast V_{\alpha }$ ($\alpha = 1,\cdots ,d$) in general do not satisfy the appropriate regularity conditions so that the flow $h_\ast g_{s,t}(\omega )$ in general does not have the same regularities as those of the flow $g_{s,t}(\omega )$.

Now for every vector field $V$ on $S^7_s$, the induced vector field $h_\ast V$, whenever it exists, is given by
\begin{equation}
\left[h_\ast V\right]^i(\gamma ^1, \cdots , \gamma ^8)=\sum _{j=1}^8\frac{\partial h^i}{\partial z^j}V^j(h^{-1}(\gamma ^1,\cdots ,\gamma ^8)),
\end{equation}
for every $(\gamma ^1,\cdots ,\gamma ^8)\in \Sigma ^7_{GM}$. Since
\begin{equation}
\frac{\partial h^i}{\partial z^j}=\sum _{l=1}^8\frac{\partial [D^{-1}]^i}{\partial \zeta ^l}z^l\frac{\partial \beta }{\partial z^j}+\frac{\partial [D^{-1}]^i}{\partial \zeta ^j},
\end{equation}
then
\begin{eqnarray}\label{medan}
\left[h_\ast V\right]^i(\gamma ^1, \cdots , \gamma ^8) & = &\sum _{j,l=1}^8\frac{\partial [D^{-1}]^i}{\partial \zeta ^l}z^l\frac{\partial \beta }{\partial z^j}V^j(h^{-1}(\gamma ^1,\cdots ,\gamma ^8))+\sum _{j=1}^8\frac{\partial [D^{-1}]^i}{\partial \zeta ^j}V^j(h^{-1}(\gamma ^1,\cdots ,\gamma ^8)).
\end{eqnarray}
The factor $\partial \beta /\partial z^j$ in the right hand side of equation \eqref{medan} may not be defined because the function $\beta $ may not be $C^1$-differentiable.
On the other hand, it is clear that
\begin{eqnarray}\label{pullback}
\frac{\partial z^j}{\partial \gamma ^i}=
\frac{\partial [h^{-1}]^j}{\partial \gamma ^i}&=&-\frac{1}{2}|D(\gamma ^1,\cdots ,\gamma ^8)|^{-3}\frac{\partial |D(\gamma ^1,\cdots ,\gamma ^8)|^2}{\partial \gamma ^i}[D(\gamma ^1,\cdots ,\gamma ^8)]^j+|D(\gamma ^1,\cdots ,\gamma ^8)|^{-1}\nonumber\\
&&\qquad \frac{\partial [D(\gamma ^1,\cdots ,\gamma ^8)]^j}{\partial \gamma ^i},
\end{eqnarray}
for every $z^j$, and is differentible on $\Sigma ^7_{GM}$.
Assuming that the function $\beta $ is $C^1$-differentiable concerning $z^i$, it can be shown through direct inspection or calculation that the vector field $h_\ast V$ is differentiable on $\Sigma ^7_{GM}$ for every differentiable vector field $V$ on $S^7_s$.

Therefore, {\em whenever} the function $\beta $ is $C^1$-differentiable on $S^7_s$, the stochastic differential equation \eqref{sde13} with differentiable vector fields $V_0$ and $V_\alpha $ ($\alpha =1,\cdots d$) corresponds to the stochastic differential equation \eqref{sdeh} with differentiable vector fields $h_\ast V_0$ and $h_\ast V_\alpha $ ($\alpha =1,\cdots d$). Thus, the flow $g_{s,t}(\omega )$ on $S^7_s$ and the corresponding flow $h_\ast g_{s,t}(\omega )$ on $\Sigma ^7_{GM}$ constructed in \eqref{inducedflow} have the same regularities. There is no difference betwen the appearence of the stochastic flow on the seven sphere viewed in standard differential structure and the appearence of the same stochastic flow viewed in the Gromoll-Meyer (exotic) differential structure.

Furthermore, since the inverse mapping $h^{-1}$ is differentiable on $\Sigma ^7_{GM}$ (see equation \eqref{pullback}), the Riemannian metric tensor $[f^{-1}]^\ast G$ on $\Sigma ^7_{GM}$, i.e. the pull-back of the Riemannian metric tensor $G$ on the standard sphere $S^7_s$, is also differentiable. This fact implies, for instance, the fact that the Fokker-Planck equation associated to the stochastic differential equation \eqref{sdeh} and the Fokker-Planck equation associated to the stochastic differential equation \eqref{sde13} have the same regularities provided that the function $\beta $ is $C^1$-differentiable. Therefore both differential structures on $S^7$ give the same description of the dynamics of the distribution function of the stochastic process understudy on seven spheres.

The differentiability of the metric tensor $[f^{-1}]^\ast G$ on $\Sigma ^7_{GM}$ implies also to the regularities of Riemannian measure on $\Sigma ^7_{GM}$ so that the entropy 
\begin{eqnarray}S[(h\ast p)(\mathbf{\gamma }|\mathbf{\gamma}';t)]=
-\int _{\Sigma ^7_{GM}}(h^\ast p)(\mathbf{\gamma }|\mathbf{\gamma}';t) \log [(h^\ast p)(\mathbb{\gamma }|\mathbf{\gamma}';t)]
|[h^{-1}]^\ast G (\gamma )|^{1/2}d\mathbf{\gamma },
\end{eqnarray}
for every $\gamma '\in \Sigma ^7_{GM}$, of the distribution function $h^\ast p=p\circ h$ on $\Sigma ^7_{GM}$ is equal to the entropy of the corresponding distribution $p$ on $S^7_s$.

\end{document}